\documentstyle[twoside,fleqn,espcrc2,epsfig]{article}

\newcommand{\BE}{\begin{equation}}
\newcommand{\EE}{\end{equation}}
\newcommand{\BEA}{\begin{eqnarray}}
\newcommand{\EEA}{\end{eqnarray}}

\newcommand{\pbp}{\bar\psi\psi}

\title{Critical Behavior at the Chiral Phase Transition
\thanks{Presented by C.~DeTar.  
        Supported by the US DOE and NSF.  
        Computations were done at SDSC, NCSA, PSC, Cornell CTC, and 
        on the UCSB Origin 2000.}}
\author{ C.~Bernard\address{Department of Physics, Washington University, St.~Louis, MO 63130, USA},
T.~Blum\address{Physics Department, Brookhaven National Laboratory, Upton, NY 11973, USA},
C.E.~DeTar\address{Physics Department, University of Utah, Salt Lake City, UT
  84112, USA
  \\ and Zentrum f\"ur interdisziplin\"are Forschung, Universit\"at Bielefeld,
   Bielefeld, Germany
},
Steven~Gottlieb\address{Department of Physics, Indiana University, Bloomington, IN 47405, USA},
U.M.~Heller\address{SCRI, Florida State University, Tallahassee, FL 32306-4130, USA},
J.E.~Hetrick$\,\null^{\rm a}$,
Beat Jegerlehner$\,\null^{\rm d}$,
K.~Rummukainen\address{Fakult\"at f\"ur Physik, Universit\"at Bielefeld,
  Bielefeld, Germany},
R.L.~Sugar\address{Department of Physics, University of California, Santa Barbara, CA 93106, USA},
D.~Toussaint\address{Department of Physics, University of Arizona, Tucson, AZ
  85721, USA 
  \\ and Center for Computational Physics, University of Tsukuba, Ibaraki 305,
  Japan
}, 
and M.~Wingate\address{Department of Physics, University of Colorado, Boulder, CO 80309, USA},
} 
\begin{document}


\maketitle
\section{INTRODUCTION}
It is generally expected that two-flavor QCD undergoes a high
temperature chiral symmetry restoring phase transition at zero quark
mass, with O(4) critical behavior\cite{PW}. Verifying this expectation
is important for understanding the phenomenology of the transition and
for facilitating an extrapolation of simulation data to physical quark
masses.  Lattice QCD at fixed $N_t$ with staggered fermions is similarly
expected to exhibit at least O(2) universality, with O(4) emerging in
the continuum limit.  For present purposes we do not distinguish O(2),
O(4), or mean field critical behavior.

One test of universality compares critical exponents.  Comparing the
critical scaling function gives further insight.  We use the standard
correspondence between QCD variables and O(N) spin variables, which
identifies quark mass $m_q/T$ with magnetic field $h$, inverse gauge
coupling $6/g^2$ with temperature $T/T_c(0)$, and chiral condensate
$\bar \psi \psi$ with magnetization $M$. A critical point is expected
to occur at zero quark mass and nonzero
coupling $6/g^2(0)$.  For studies at fixed $N_t$ we use\cite{KL}
\begin{eqnarray*}
  h &=& am_q N_t \cr
  t &=& T/T_c(0) - 1 = 6/g^2 - 6/g^2(0)
\end{eqnarray*}

Critical scaling theory predicts that for small quark masses
we have the scaling relation \cite{Amit}
\begin{displaymath}
  \bar \psi \psi h^{-1/\delta} = f_{QCD}(x) = c_y f_{O(4)}(c_x x)
\end{displaymath}
where  $x = t h^{-1/\beta\delta}$, $c_x$ and $c_y$ are scale constants,
$f_{QCD}(x)$ is the critical scaling function for QCD and $f_{O(4)}(x)$ is
that for O(4).  Not shown are nonleading nonscaling 
contributions analytic in $t$ and $h$.

\section{A CRITICAL EXPONENT}

We extended an old data sample at $N_t = 4$, which was done with the standard
one-plaquette gauge action plus two-flavor staggered fermion action.  Our new
sample decreases the quark mass to $am_q = 0.008$ and increases the spatial
lattice size $L$ to 24. We also reanalyze old data at $N_t = 6,8,12$
\cite{Nt12}.  The latter all have $L = 2 N_t$.

The mixed plaquette/chiral condensate susceptibility 
$$
\chi_{mt} = \partial (\bar \psi \psi)/\partial (6/g^2)
$$ 
has a peak, which can be used to locate the crossover.  
We extrapolate the peak height to infinite volume as shown below.

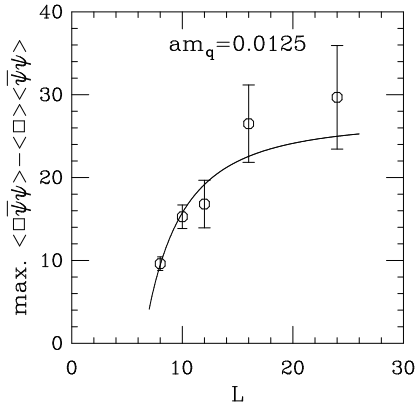
\vspace*{3mm}
\centerline{\epsfig{file=pbpA_ext_0125.ps,width=60mm}} 

The infinite volume value should scale as
$$
  \chi_{mt,\rm max} \sim (am_q)^{(\beta - 1)/\beta\delta} 
    \approx (am_q)^{-0.33}
$$
The figure below shows a strong departure from scaling at present
quark masses, in agreement with similar findings of the JLQCD and
Bielefeld groups for a variety of susceptibilities\cite{U96}.

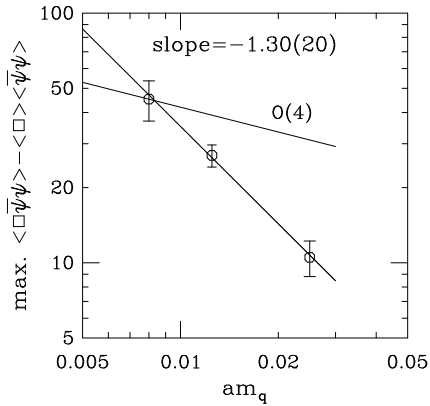
\vspace*{3mm}
\centerline{\epsfig{file=pbpA_ext.ps,width=60mm}} 

\section{SCALING FUNCTION AT $N_t = 4$}

To find the critical coupling at zero quark mass, $6/g^2(0)$, we use
the scaling relation for the crossover (pseudocritical) coupling \cite{KL}:
$$
  6/g^2_{pc} = 6/g^2(0) + x_{pc} (N_t am_q)^{1/\beta\delta}
$$
which works well.

Neglecting nonscaling contributions to $\pbp$, we construct the
QCD critical scaling function (for $x>0$) using O(4) critical
exponents \cite{KK} and compare with the scaling function for O(4)
\cite{O4} in the figure below. Vertical and horizontal displacements
of the log-log scaling curves are permitted and correspond to
adjusting $c_x$ and $c_y$.


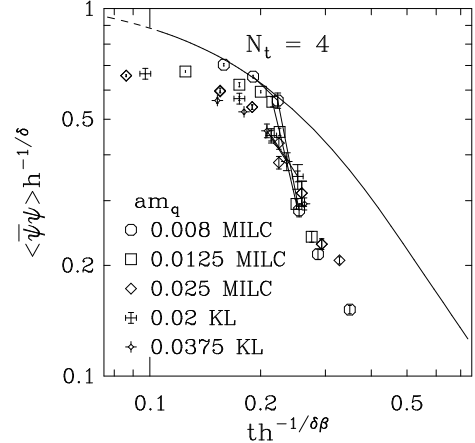
\centerline{\epsfig{file=pbpnt4.change.5225.ps,width=65mm}} 
The line segments locate the crossover in the QCD data and the dashed line on
the O(4) curve, the crossover in O(4).  The newer data are plotted with 
octagons and squares.

We see: (1) The new data show generally worse agreement with the O(4)
scaling curve. (2) In terms of these scaling variables and at these
quark masses, the O(4) critical region is quite small, if it is to be
found at all.  Equivalently, nonnegligible nonscaling contributions may
be masking critical behavior in this regime.  (3) Nonscaling contributions
could be displacing the crossover location.

\section{LARGER $N_t$}

A similar analysis at larger $N_t$ is shown below.  Perhaps there is
improvement with increasing $N_t$.  However, our $N_t = 12$ sample
includes data only at a single quark mass, making it our weakest
example.  Furthermore, for $N_t > 4$ we have no results for $L > 2
N_t$.

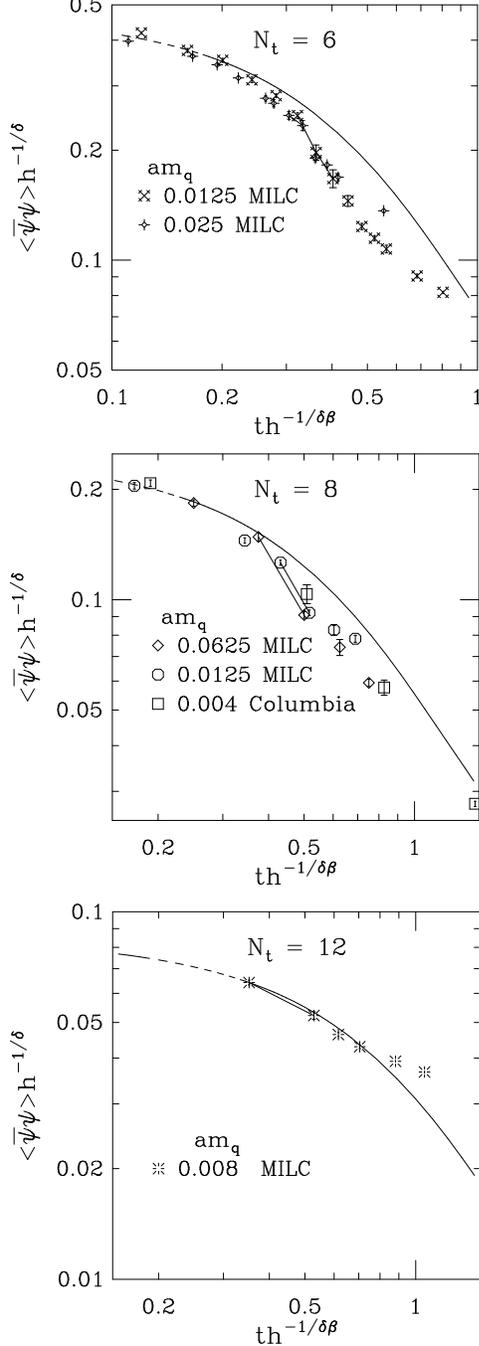
\centerline{\epsfig{file=pbpnt6.533.ps,width=65mm}} 
\centerline{\epsfig{file=pbpnt8.540.ps,width=65mm}} 
\centerline{\epsfig{file=pbpnt12.555.ps,width=65mm}} 

\section{SPECULATIONS}

New results at smaller quark mass and larger volume at $N_t = 4$ have
raised doubts about the extent of the previously observed agreement
between QCD and O(4) \cite{KL,Nt12}.  In striking contrast with Wilson
fermions with an improved gauge action \cite{Wscale}, the conventional
staggered fermion action with the conventional choice of scaling
variables does not show good agreement with the O(4) scaling function
at present quark masses and temperatures.

Why? (1) Perhaps the critical region is attained only when $\pi$ and
$\sigma$ correlation lengths are considerably greater than $1/T_c$.
Here, typically, they are smaller. (2) At strong coupling the
variables $6/g^2$ and $ma$ may have a strongly nonlinear
correspondence with spin model temperature and magnetic field $T, h$,
which becomes approximately linear only in a small region around the
critical point.  The nonlinearities may become weaker in the weak
coupling limit and perhaps may be diminished with improved actions.
Perhaps with a more suitable choice of scaling variables, we may find
better agreement.


\end{document}